\documentclass[pdflatex,sn-mathphys-num]{sn-jnl}

\usepackage{geometry}[a4paper,left=1.3cm,right=1.3cm,top=1.5cm,bottom=1.5cm]
\usepackage{multirow}%
\usepackage{amsmath,amssymb,amsfonts}%
\usepackage{amsthm}%
\usepackage{mathrsfs}%
\usepackage[title]{appendix}%
\usepackage{xcolor}%
\usepackage{textcomp}%
\usepackage{manyfoot}%
\usepackage{booktabs}%
\usepackage{algorithm}%
\usepackage{algorithmicx}%
\usepackage{algpseudocode}%
\usepackage{listings}%

\theoremstyle{thmstyleone}%
\theoremstyle{thmstyletwo}%

\theoremstyle{thmstylethree}%

\usepackage{graphicx}%

\begin{document}
\newgeometry{
  left=2cm,right=2cm,top=2cm,bottom=2cm,
  headheight=0pt,headsep=0pt,footskip=0pt
}
\pagestyle{plain}

\title{Assessment of a Hybrid Energy System for Reliable and Sustainable Power Supply to Boru Meda Hospital in Ethiopia}

\author*[1]{\fnm{Tegenu} \sur{Argaw Woldegiyorgis}} \email{argaw2009@gmail.com}
\author[2]{\fnm{Hong} \sur{Xian Li}} 
\author[3]{\fnm{Fekadu} \sur{Chekol Admassu}}
\author[1]{\fnm{Merkebu} \sur{Gezahegne}}
    \author[1]{\fnm{Abdurohman} \sur{Kebede}}
    \author[1]{\fnm{Tadese } \sur{Abera}}
    \author[4]{Haris Ishaq}
     \author[1]{\fnm{Eninges} \sur{Asmare}}

\affil[1]{\orgdiv{Department of Physics, College of Natural Sciences}, \orgname{Wollo University},  \city{Dessie}, \postcode{1145}, \country{Ethiopia}}
\affil[2]{\orgdiv{Faculty of Science, Engineering \& Built Environment}, \orgname{Deakin University}, \country{Australia}}
\affil[3]{\orgdiv{Department of Chemistry, College of Natural Sciences}, \orgname{Wollo University},  \city{Dessie}, \postcode{1145}, \country{Ethiopia}}
\affil[4]{\orgdiv{Department of Chemical Engineering}, \orgname{King Fahd University of Petroleum \& Minerals}, \city{Dhahran}, \postcode{31261}, \country{Saudi Arabia}}


\abstract{This study aims to evaluate the techno-economic feasibility of hybrid energy systems (HES) including Grid for providing reliable and sustainable power to Boru Meda Hospital, Ethiopia. HOMER pro 3.11.2 was used to design and evaluate a novel, integrated optimization and comparative assessment of diverse HRES, specifically adjusted to the energy consumptions and available resources of the Hospital. The scenario evaluation showed that interconnecting photovoltaic (PV), biomass generator (BG), wind power (WP), diesel generator (DG), battery, and converter can effectively provide the Hospital's daily energy consumption of 11,214.66 kWh while conforming reliability and reducing emissions. The PV/BG/batt/conv configuration emerged as the most cost-effective and sustainable alternative, attaining the lowest LCOE of \$0.339/kWh, an NPC of \$25.7 million, and a 100\% renewable energy fraction with simple pay back of 7.26 yr. As a result, the operational cost associated with the consumption of 500.00 L of diesel per month can be entirely avoided. The DG-integrated hybrids exhibit advanced techno-economic capability with significant worth, strong ROI (20\%) and IRR (18\%), endorsed by fast capital recovery (7.21-8.71 years). Overall, the hybrid system offers an optimal balance of cost, reliability, and sustainability, making it a promising and scalable solution for electrification of energy scare institution and areas in Ethiopia, thereby contributing to national sustainable energy development goals.}

\keywords{Available resources, efficient energy production, energy demand, emission, environment friendly, techno-economic, power shortage}



\maketitle

\vspace{-10pt} 
\begin{table}[h!] 
    \centering
    \caption{Summary of Key Nomenclature}\label{Tab1}
    \begin{tabular}{ll|ll} 
        \hline
        \textbf{Abbrev.} & \textbf{Extended form} & \textbf{Abbrev.} & \textbf{Extended form} \\ \hline
        BG & Biomass Generator & CRF & Capital Recovery Factor \\
        DC & Direct Current & DG & Diesel Generator \\
        $D_{PV}$ & Parameter of Panel Performance & $E_{dl}$ & Total Daily Load \\
        GHG & Greenhouse Gases & HES & Hybrid Energy System \\
        HRESs & Hybrid Renewable Energy Systems & IRR & Internal Rate of Return \\
        kJ/kg & Kilojoule per Kilogram & kWh & Kilowatt Hour \\
        kWh/day & Kilowatt Hour per Day & LCOE & Levelized Cost of Energy \\
        MRP & Minimum Renewable Penetration & NPC & Net Present Cost \\
        $P_{WP}$ & Turbine Power & $P_{inv}$ & Power Inverter \\
        $P_{peak}$ & Peak Load Power & $P_{R}$ & Derating Factor \\
        $P_{r}$ & Rated Power & T$_{cell}$ & Temperature of Solar Cell \\
        WP & Wind Power & $v_{cut-in}$ & Cut-in Velocity \\
        $v_{cut-out}$ & Cut-out Velocity & $v_r$ & Rated Velocity \\
        \hline
    \end{tabular}
\end{table}

\section{Introduction}

 The dependence on traditional energy sources for electricity generation creates serious environmental, economic, and logistical challenges, particularly in low income countries \citep{arevalo2025decarbonizing,ochoa2025pathways}. Emissions from fossil fuel combustion have had a significant impact on climate \cite{shahveran2025replacing}, contributing to global warming and threatening both human health and environmental safety \cite{POPOOLA2025100290}. The power sector, as a of the leading emitters and one of the most climate-vulnerable industries, therefore, must advance rapidly toward decarbonization \cite{obiora2024assessing}. Mitigating these impacts requires the shift to renewable energy, which is essential for decarbonization \cite{raza2025carbon} and to achieve overall sustainability objectives \cite{raineri2025power,gayen2024review}. Clean energy plays a vital role not only in reducing emissions \cite{aridi2025eco}, but also in driving economic growth and fostering technological innovation in modern societies \cite{mandel2024towards}. Recently, IEA unveiled a comprehensive road map describing strategies for the global energy sector to achieve net-zero emissions by 2050 \cite{bera2025advancing}. \\
 
 However, many sub-Saharan African countries continue to face limited energy access \cite{mossisa2025transitioning} and volatile global energy prices, often resorting to subsidies to make energy affordable for low-income households \cite{sibanda2024animal}. The region faces a complex challenge marked by widespread energy poverty \cite{quarcoo2025revisiting}, increased climate related risks, and limited development progress \cite{tomala2021towards}. Ensuring reliable and clean energy in sub-Saharan Africa requires stepping beyond dependence on wood and fossil fuels, which is crucial to improving access to electricity and safe cooking energy \cite{lahnaoui2024assessing}. The worldwide effort to achieve net-zero emissions by 2050 underscores the pressing need to shift toward clean energy systems based on renewable resources and supported by low-emission technologies \cite{boafo2025energy}. Recent investigations have emphasized the growing importance of HRES. For example, it can effectively reduce power outages, increase clean energy utilization, and reduce GHGs emissions \cite{karapidakis2024zero}. In the study by \cite{panbechi2025enhancing}, HRES represents a promising approach to reduce carbon emissions, improve energy security, and ensure a stable and reliable electricity supply. 
  In \cite {habib2025multi}, a PV, BG and converter system delivered electricity at a COE of \$0.041/kWh, cost \$1.89M, cut 1,220 tone a of $CO_{2}$ annually and operated fully on clean energy. Similarly, another study using HOMER software identified an optimal configuration that combine wind, diesel and battery systems, generating electricity at a COE of \$0.213/kWh and a total cost of \$628,571, where the inclusion of batteries significantly reduced the use of diesel generators usage \cite {alanazi2025optimal}. Furthermore, \cite{roy2025techno} reported that the PV/WT/bat/conv/BG system produced the least expensive electricity (\$0.278/kWh, \$1.61M total cost), whereas the bat/conv/BG system was the most costly (\$0.455/kWh, \$2.63 million), providing valuable information to design reliable off-grid hybrid energy systems. The  investigation by \cite{ishraque2025solar}, shows that the optimal system, using 89.1\% clean energy, sells 192,161 kWh per year to the grid, produces electricity at \$0.0132/kWh, provides a return of 73\% with a 1.4-year payback, and emits very little $CO_{2}$ (18,647 kg) and $SO_{2}$ (80.8 kg) annually, demonstrating its practicality and scalability for clean stable power. Using HOMER, a PV/diesel/battery system a poultry farm in North Carolina shows that a 20 kW grid-tied PV can save \$3,200, batteries are still costly, off-grid systems cost \$370,000–\$560,000 more, and battery, PV system, and diesel prices are the most affected cost \cite {wang2025techno}.\\

  A recent study shows major power gaps in SSA health centers and finds that standalone PV can provide clean, low-cost power, reducing the travel time of 281 million people by approximately 50 minutes \cite {moner2021achieving,soto2022solar,alsagri2021techno}. Investigation of \cite {alrbai2023sustainable} shows that the off-grid wind-biogas system provides stable, clean, and low-cost power to remote health centers, promoting sustainable and eco-friendly energy use. Grid/PV is the cheapest for grid systems (\$282,492 NPC, \$0.0401/kWh, 55\% $CO_{2}$ cut), while off-grid generator/PV/battery costs \$1.19 M, \$0.342/kWh, cutting 65\% $CO_{2}$; full renewable systems achieve zero emissions but are too costly \cite{caglayan2025optimization}. In rural Ethiopia, the use of growing, saving 43.68 L of kerosene and 107 kg of $CO_{2}$ per home yearly \cite{wassie2021socio}. The study by \cite{salau2024design} found that the PV/diesel/ battery system for Shinshicho Hospital, Ethiopia, reaches a load of 276 kWh/day at \$0.187/kWh, at a cost of \$160,500, proving it economical and sustainable. As designed in HOMER Pro 3.14.2, an off-grid PV-wind microgrid with battery storage costs \$136,082, produces power at \$0.0688/kWh, does not emit carbon and is fully renewable \cite{ali2025sustainable}. For off-grid rural electrification in Ethiopia, a PV-wind-battery-diesel hybrid, modeled in HOMER Pro, is the cheapest, has a peak of 19.6 kW at \$0.207/kWh, costs \$82,734, and cuts 37.3 tons $CO_{2}$ per year \cite {gebrehiwot2019optimization}. In Kiltu village, Ethiopia, a PV-diesel/ battery system is the most efficient, costing \$0.326/kWh with \$1.42 M NPC, cutting 49\% $CO_{2}$, meeting seasonal farm needs and improving health, energy resilience and rural development \cite {woldegiyorgis2025techno}. In the study of \cite {benti2022techno}, the PV/DG/battery system stands out as the most feasible and optimized option, minimizing costs while maximizing environmental benefits and demonstrating superior economic viability and sustainability.\\

The recent study evaluates the transition from diesel to PV/diesel hybrid systems, revealing a 22\% cost reduction for users, a 7.6-year payback period, a profit of USD 44,698, and an annual $CO_{2}$ reduction of 8.26 tons, confirming the economic and environmental suitability of the system for rural electrification \cite{omar2024green}.
Frequent power outages in developing countries often result from an imbalance between electricity supply and demand, severely affecting the healthcare sector \cite{soto2022solar}.
Research in \cite {adedoja2024techno} shows that custom-made hybrid renewable systems in rural health centers help improve health care and access to clean energy. Nevertheless, many studies in sub-Sahara Africa and Ethiopia have assessed the feasibility of a stand-alone or hybrid system for Hospitals, farms, schools, and rural communities, but few have provided a comprehensive evaluation of multiple hybrid designs. This study uses HOMER Pro to design an optimized, reliable and sustainable hybrid energy design for Boru Meda Hospital, Ethiopia. This investigation presents a novel hybrid optimization and comparative evaluation of various HRES using HOMER pro 3.11.2, specifically adjusted to the energy consumptions and available resources of Boru Meda Hospital, Ethiopia.

The remainder of this article is organized as follows: chapter two describes the methodology and data collection techniques, chapter three presents and critically discusses the results, and chapter four summarizes the main findings and provides the collusion.
\section{Method and Material}

\subsection {Description of the Study Area}
 \begin{figure}[ht]
      \centering
      \includegraphics[scale = 0.45]{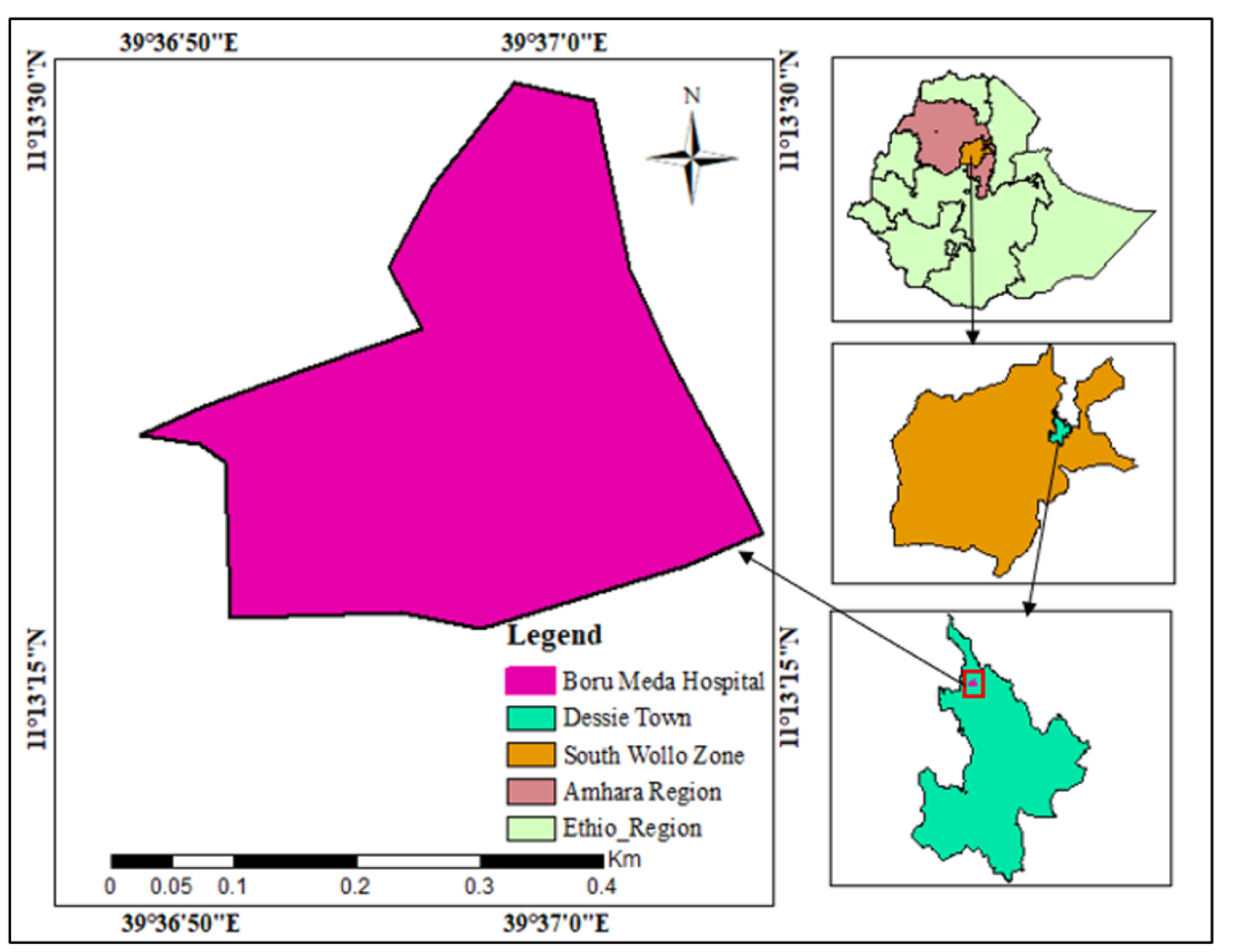}
      \caption{Map of the study area }\label{Map} 
   \end{figure}
Boru Meda is located approximately 10 km along the main road to the Gishen Mariam Church in the South Wollo Zone of the Amhara Regional State, Ethiopia. It is a key center for dairy production and mixed crop livestock farming, supplying fresh milk to Dessie, which is the capital city of South Wollo and the surrounding markets. Livestock is a central living hood in the house, providing milk, organic fertilizer, and income. They produce approximately 3494798.32 kg/yr of animal waste (dung) from various species including oxen, cows, heifers, bulls, horses, calves, donkeys, mules, sheep, goats, and chickens.
 
\begin{figure}[ht]
      \centering
      \includegraphics[scale = 0.40]{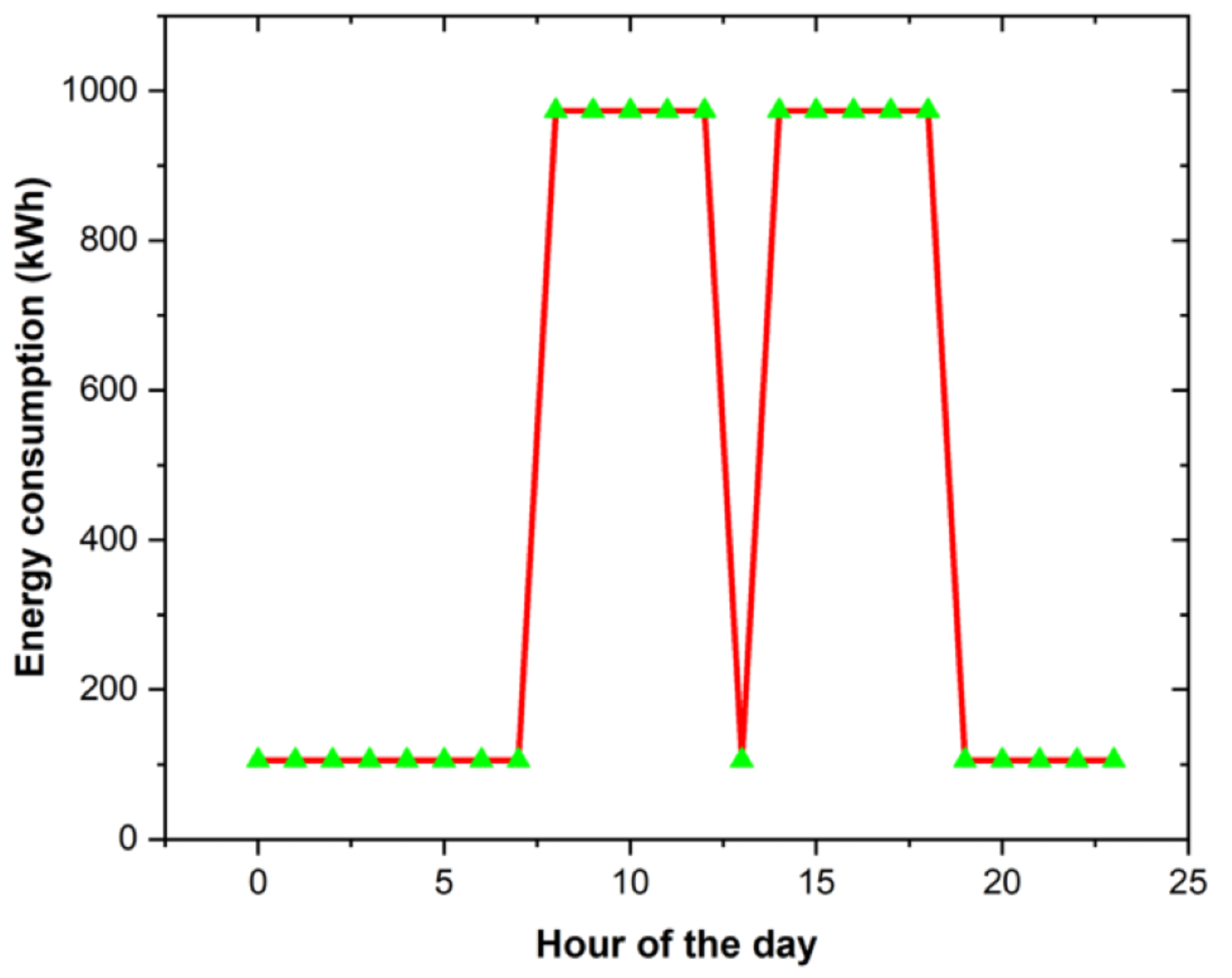}
      \caption{Hourly distribution of energy consumption at Boru Meda Hospital}\label{Hour} 
   \end{figure}
   
    \begin{figure}[ht]
    \centering
     \includegraphics[scale = 0.40]{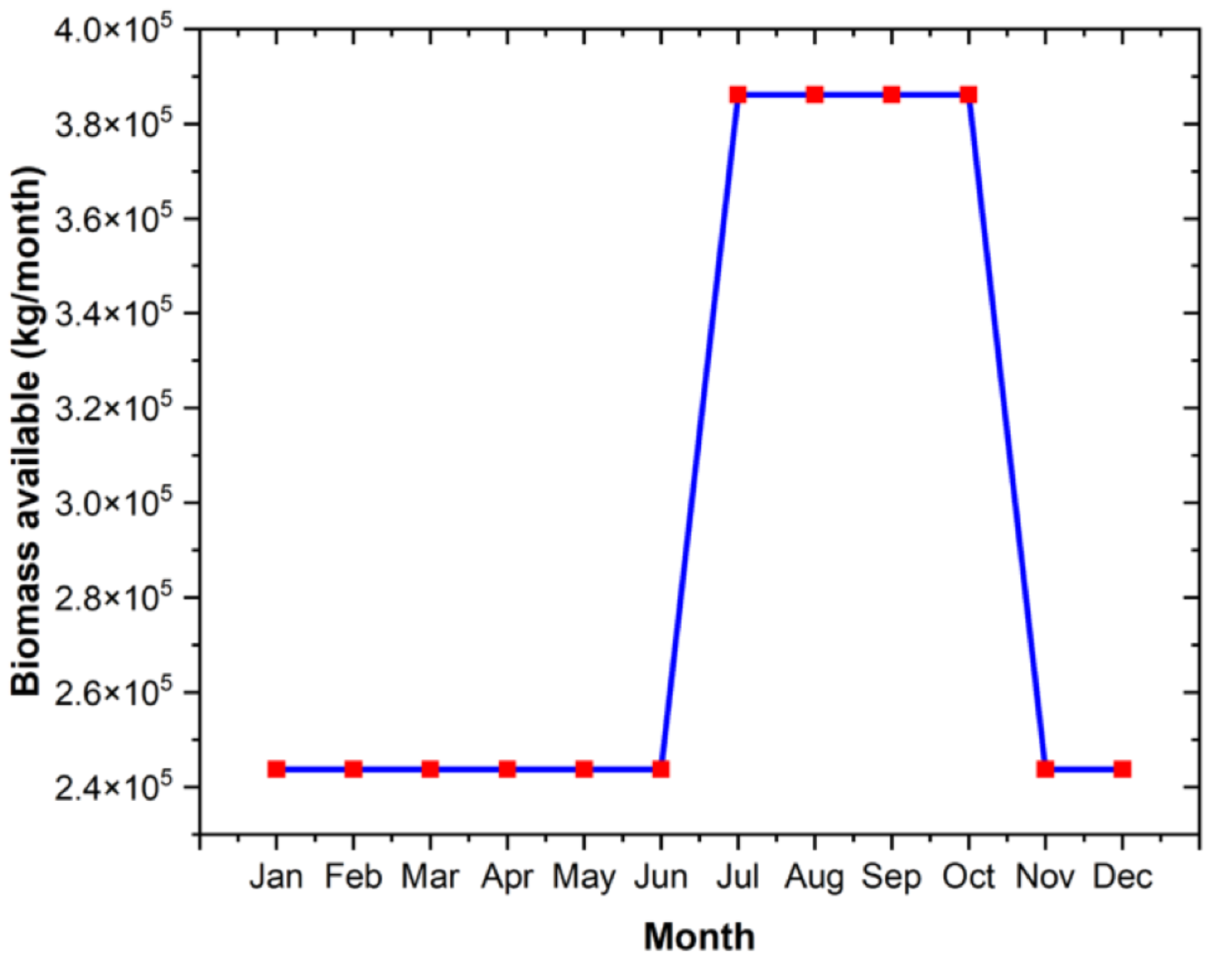}
      \caption{Monthly variation of available biomass at Boru Meda Hospital}\label{Month} 
   \end{figure}
   
Boru Meda Hospital, a major regional healthcare facility, is located at $11^{0}$$13^{'}$$22^{"}$ N latitude and $39^{0}$$36^{'}$$58^{"}$ E longitude, as shown in Fig.  \ref{Map}. provides a wide range of clinical and specialized healthcare services, including ophthalmology, orthopedics, comprehensive eye care and rehabilitation support, physiotherapy, general and specialized surgical procedures, along with the diagnosis and treatment of infectious diseases such as HIV/AIDS and tuberculosis (TB). However, the hospital often experiences power shortages, power outages, and poor electricity supply. Despite these challenges, the area has abundant solar and biomass resources, along with moderate wind potential, offering significant opportunities for the installation of hybrid renewable energy systems.

\subsection{Data collection}
   
 \begin{figure}[ht]
      \centering
      \includegraphics[scale = 0.40]{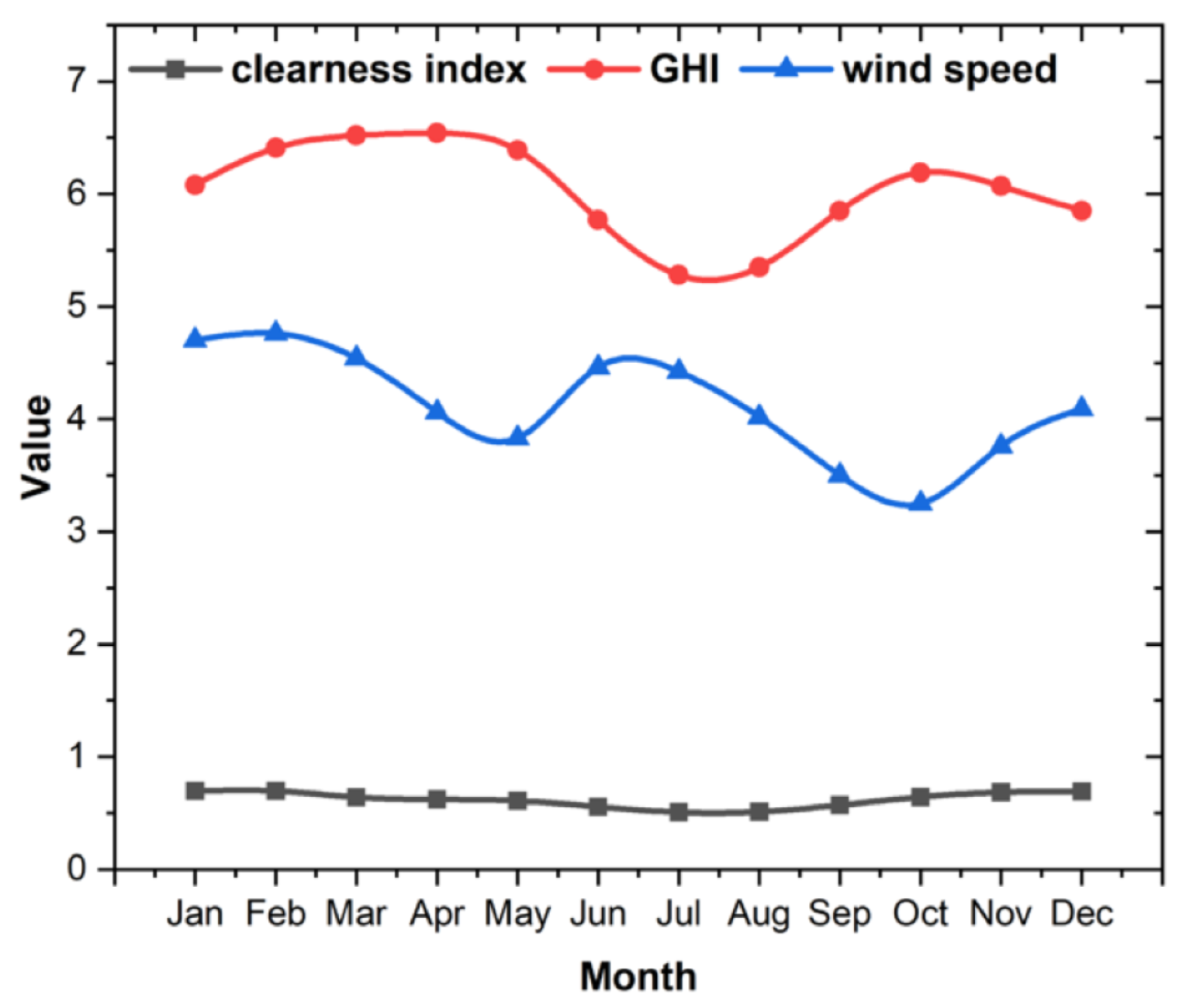}
      \caption{GHI, wind speed, and clearness index at Boru Meda Hospital}\label{GHI} 
   \end{figure}
   
      \begin{figure}[ht]
      \centering
      \includegraphics[scale = 0.40]{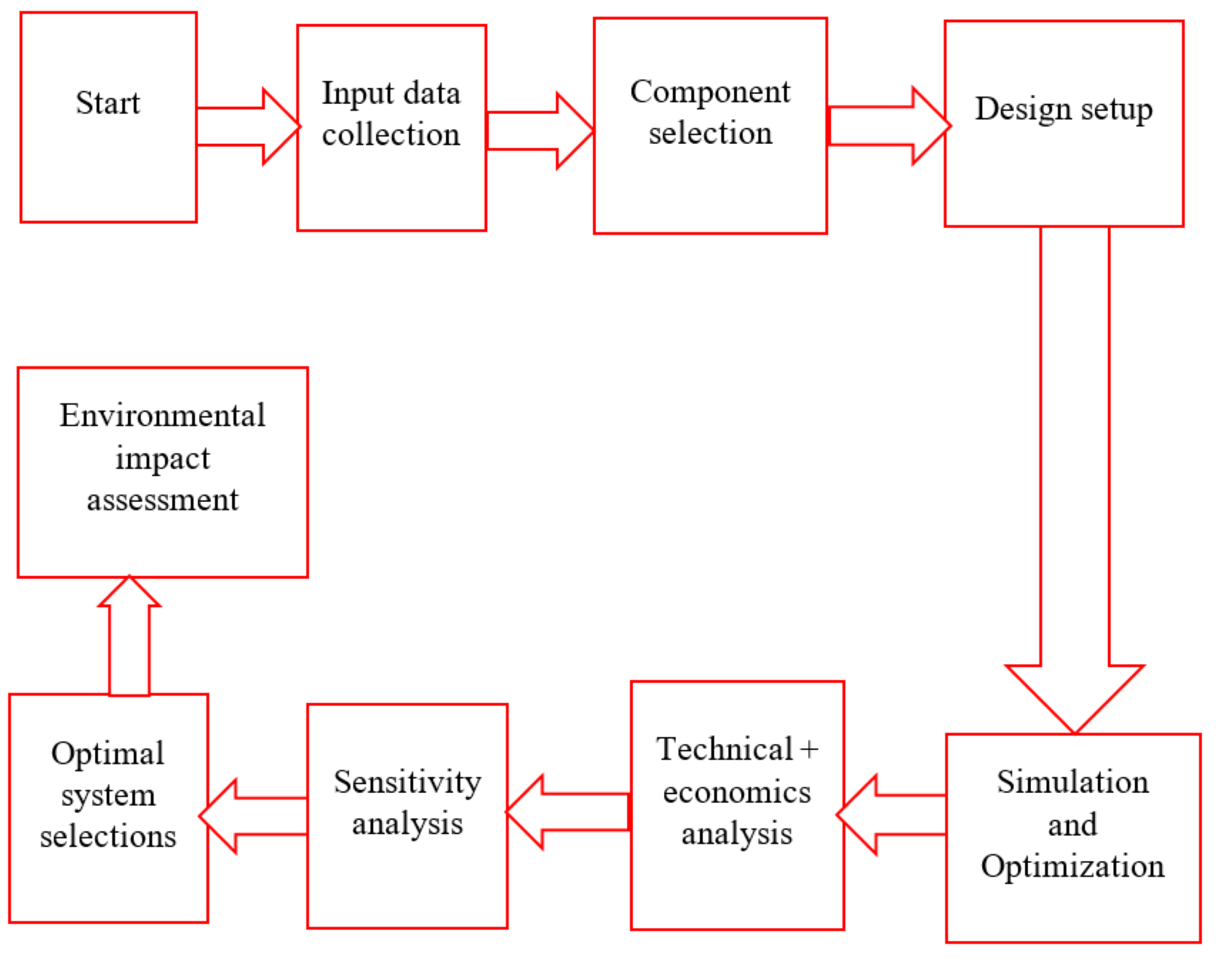}
      \caption{Optimal system design and selection process}\label{Opt} 
   \end{figure}
   
Assessing available resources and energy consumption is crucial to mitigate the power shortage of the Hospital and to replace fossil fuels (diesel). 
The energy consumption patterns of various services that incorporate residential areas, lighting systems, and medical equipment from different departments were  calculated comprehensively to design a hybrid system. The total energy requirement was calculated using the following equation \cite{woldegiyorgis2025techno}: 
\begin{equation}
E_{dl}(\text{kWh}) = \frac{1}{100} \left( \sum_{i=1}^{n} A_{n} P_{n} H_{n} \right)
\label{eq:energy_demand}
\end{equation}
Where $E_{dl}$ (KWh) - the total daily load for a specific service type,
$A_{n}$ - the number of appliances within that service type, $P_{n}$ - the rated power of each appliance and $H_{n}$ - the total hours of operation. 

Fig. \ref{Hour} illustrates the 24-hour energy consumption pattern of the Hospital, showing a distinct diurnal variation corresponding to the operational schedule. During  nighttime hours (00:00 to 07:00), energy demand remains low at approximately 105.85 kWh/hr, sustained primarily by essential systems such as emergency lighting, refrigeration, and life-support equipment. A sharp increase in consumption occurs between 08:00 and 12:00, reaching a peak of 973.28 kWh/hr as clinical, diagnostic, and administrative activities reach full operation. Around 13:00, a temporary decline in demand reflects reduced activity during the midday break, followed by a renewed rise between 14:00 and 18:00 with the continuation of afternoon services. After 18:00, energy consumption progressively decreases and returns to 105.85 kWh/hr and maintains stability throughout the late evening and night (19:00–23:00).

Fig. \ref{Month}, illustrates the monthly fluctuation in the available biomass (kg/day) of the Hospital. The monthly average availability varies between $2.4 * 10 ^{5}$ kg/day and $3.9 * 10^{5}$ kg/day due to the variation in the season of agricultural activities and livestock management practices.
   
During the months (November – June), biomass availability remains relatively stable at approximately $2.4 * 10^{5}$ kg/day, corresponding to a period of limited crop residues and animal dung production due to reduced feed supply. Biomass is at its peak between July and October at approximately $3.9 * 10^{5}$ kg/ due to the rainy and post-harvest seasons. The July–October period represents the peak of biogas production and hybrid renewable energy applications, while the dry season emphasizes the importance of effective biomass storage and resource management strategies to maintain a reliable energy supply throughout the year.

 Fig. \ref{GHI} presents the monthly variation of GHI, the wind speed and the clearness index of the study site, which were taken from NASA. The GHI varies between 5.2 and 6.5 $kWh/m^{2}$/day, showing clear seasonal patterns. Higher values occur from January to May and October to December, associated with clear skies and longer sunshine hours, while lower values from June to September correspond to the duration of rain with increased cloud cover. Despite this decrease, solar radiation remains adequate for effective energy production throughout the year. Wind speed varies between 3.0 and 4.8 m/s, peaking from January to March and decreasing between August and October. In addition, the wind speed is low in May and November. The inverse relationship between wind speed and solar radiation indicates a complementary pattern, which increases the stability of hybrid energy generation. The clearness index remains steady between 0.45 and 0.55, reflecting moderate atmospheric transparency and confirming the seasonal impact of cloud cover.
 
The presented Fig. \ref{Opt} shows a well-structured and methodical framework for the design and evaluation of a hybrid energy system tailored for Boru Meda Hospital. The process begins with data collection and component selection, and then  subsequently advances through critical phases such as system configuration, environmental impact assessment, and optimal system determination. This systematic sequence reflects a comprehensive and integrative approach that encompasses technical, economic, and environmental aspects to achieve a balanced and sustainable design outcome.
   
Notably, the inclusion of stages like sensitivity analysis, simulation, and optimization reveals the iterative and data based nature of the methodology, enabling continuous refinement of system parameters to improve efficiency and adaptability under diverse operational conditions. Moreover, the integration of extensive technical and economical analysis that confirms the system performance improvements are evaluated in parallel with financial feasibility,thereby supporting the development of optimized, cost effective, and practically deployable renewable solutions.

\section{Results and Discussion}
 \subsection{Results}
In Ethiopia, particularly at the Boru Meda Hospital, unreliable electricity supply, frequent power shortages, and repeated outages are common challenges. These interruptions are currently supported by DG, which requires an average of 500.00 L per month. The average consumption of Hospital’s is approximately 11,214.66 kWh/day. In this paper, a comprehensive techno-economic assessment was carried out using HOMER pro software, in which  more than 40,000 design simulations were conducted to evaluate a wide range of possible system setups. From these designs, approximately 1000 were identified as technically feasible based on key performance indicators such as NPC, LCOE, system efficiency, and reliability.The optimized HES was designed to meet the Hospital’s electrical demand reliably and sustainably. Based on these design evaluations, an optimized hybrid energy system was proposed to confirm a reliable, sustainable, and cost-effective electricity supply capable of meeting the Hospital's energy consumption. 

A Continuous and stable power supply could be achieved based on equations (\ref{pv} to \ref{n}) using HOMER pro software and integrating PV, WP, BG, DG, battery storage and a converter with the Grid system. This hybrid design minimizes emissions, reduces operating costs, and provides resilient and scalable energy solutions for the Hospital’s facilities.

\begin{figure}[ht]
      \centering
\includegraphics[scale = 0.45]{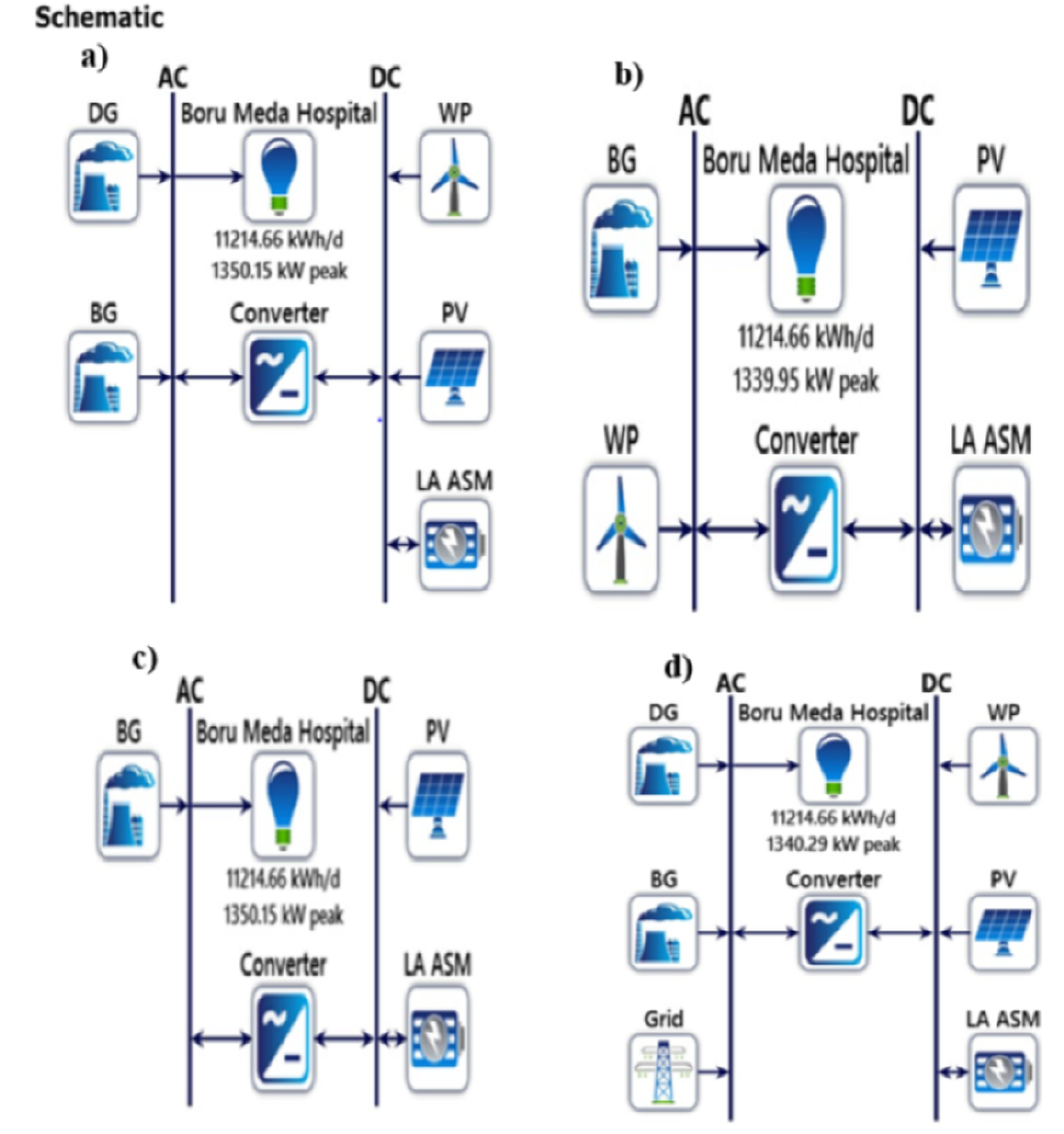}
    \caption{Hybrid Renewable energy system design for  at Boru Meda Hospital a) BG-DG-WP-PV-conv-batt b) WP-BG-PV-con-batt  c) BG-PV-conv-batt d) Grid-BG-DG-WP-PV-conv-batt}\label{scheme} 
   \end{figure}
 
  \subsubsection{ \bf {Mathematical Representation PV Systems}}
   
PV systems produce DC power by converting sunlight into electrical energy. Their output changes with solar irradiance and cell temperature. Crystalline silicon panels are the main technology, with the output given in Equation (\ref{pv}) \cite{lata2024optimization,NSAFON2020100570,OLATOMIWA2015435,PRUM2024e29369,AMMARI2022399}.

\begin{align}
     PV(t) = (P_{R}) (D_{PV})\left[\frac{GHI(t)}{GHI_{ref}}\right][1+\alpha_{P}(T_{cell}(t)-T_{ref}]\label{pv}
   \end{align}

Where $P_R$ - the derating factor, $D_{PV}$ - the panel performance ratio, and GHI(t)- the solar irradiance on the panel surface (kW/m$^2$).The reference irradiance
 $GHI_{ref}$ is 1000 W/m$^2$, $\alpha_{P}$ - the temperature coefficient and $T_{cell}$ - the solar cell temperature.

\subsubsection{\bf {Modeling of Biomass Energy Conversion Systems}}
 
Biomass has traditionally served for multiple purposes, but recent technological advancements help convert it to electricity through gasification, pyrolysis, combustion, fermentation, or anaerobic digestion, providing  efficient results \cite{lata2024optimization}. The potential for generating bioenergy and electricity from accessible manure resources is evaluated by analyzing the available feedstock supply \cite{Chowdhury2020FeasibilityAC,Nawab2023SolarIP, ALNAJJAR2022103538, islam2018thorough}. The power output of the BG is evaluated using Equation (\ref{ebm}) \cite{lata2024optimization}.

\begin{align}
  E_{BM} = (P_{BM})(CUF)(365)\left[\frac{\text{$operating$ $hours$}}{day}\right]\label{ebm}
\end{align}

Where, $P_{BM}$ - the rated capacity of the biomass gasifier and CUF -  the capacity utilization factor, taken as 0.25. The peak gasifier (kW), calculated by the available husk, is acquired using equations (\ref{pbm}) :

\begin{align}
    P_{BM} = (BM_{TA})(1000)(CV_{BM})(\eta_{BMG})\label{pbm}
  \end{align}

Where, $BM_{TA}$ - the annual availability of rice husk (tons/year), $\eta_{BMG}$ - the efficiency of the BG (\%) and $CV_{BM}$ - the calorific value of the biomass (kJ/kg). The minimum operating load ratio for the biomass system is set at 0.3 \cite{islam2018thorough}.
   
  \subsubsection{\bf {Modeling of Diesel-Based Power Generation Systems}}
  
DGs use fossil fuels for backup power; proper size reduces runtime and fuel use is calculated by equation (\ref{dsl}): \cite{lata2024optimization,Samatar2024PerformanceAO}.
   
   \begin{align}
      C_{DG} = f_1(R_{DG}) + P_{DG})f_2\label{dsl}
   \end{align}
Where $ C_{DG}$ - fuel consumption in liters, $f_1$ - the fuel curve coefficient (liters per hour), $R_{DG}$
- the rated power of the DG and $P_{DG}$ - the actual power output in kW.
   
   \subsubsection{\bf {Modeling of Wind Turbine Power Systems}}
   
The power produced by a wind turbine is affected by factors such as turbine characteristics, wind speed conditions, conversion efficiency, and maintenance practices \cite{OLATOMIWA2015435,NSAFON2020100570,PRUM2024e29369,NGILAMULUMBA2023100381}. Additional factors that impact the output are the height of the tower and the  power curve of the turbine \cite{emezirinwune2024off}. A wind turbine converts wind energy into electricity under standard conditions, as shown in Equation (\ref{pg3}) \cite{DAS2021100673,WANG2024e32712}.
   \begin{align} 
   P_{WP}(t) = \begin{cases}
               0, & v(t) \leq v_{cut-in} ~ or~ v(t) \geq v_{cut-out}  \\
               P_{r}\left[\frac{v^3(t) - v^3_{cut-in}}{v^3_{r} - v^3_{cut-in}}\right], & v_{cut-in} < v(t) < v_r \\
               P_{r}, & v_r \leq v(t) \leq v_{cut-in}\label{pg3}
            \end{cases}
   \end{align}
   
where $P_{r}$ , $v_r$,  $v_{cut-in}$ and $v_{cut-out}$ - the rated power of the turbine, the rated wind speed, the cut-in speed and the cut-out speed, respectively. Wind speed changes with altitude and measurement height. $P_{WP}$ - the power output turbine, calculated from the wind speed of the hub height as indicated in equation (\ref{al}) \cite{DAS2021100673}.
   
   \begin{align}
     v(t) = v_{ref}\left[\frac{H_{G3}}{H_{ref}}\right]^{\alpha}\label{al}
   \end{align}
The reference height ($H_{ref}$) and the hub height of the turbine ($H_{G3}$) are key factors in the evaluation of wind energy. The wind shear coefficient, known as the Hellmann exponent, is denoted by the dimensionless parameter $\alpha$, which typically varies between 0.1 and 0.4 \cite{FOSSOTAJOUO2023101107}.
   
   \subsubsection{\bf {Modeling  of Battery Storage Systems}}
   
The excess energy from renewable sources is stored in batteries to provide electricity during the low generation period \cite{OLADIGBOLU20215547}. The storage capacity required  is calculated using   equations (\ref{batt}) and (\ref{inv1}) \cite{lata2024optimization}.
   
   \begin{align}
      E_{Batt}(t) = E_{Batt}(t - 1) + E_{Excess}(t)(\eta_{Batt})(\eta_{inv}) \label{batt}
   \end{align}

   \begin{align}
      E_{Excess}(t) = \left[E_{AC}(t) + E_{DC}(t)\eta_{inv}\right] - E_{Demand}(t) \label{inv1}
   \end{align}
   
   Where $E_{Batt}$ - the energy stored in the battery at a given time, $E_{Batt}(t - 1)$ - the energy stored in the previous state, $\eta_{Batt}$ is the efficiency of the battery, $\eta_{inv}$ - the efficiency of the inverter, $E_{AC}$ - the  output energy of the inverter and $E_{DC}$ - the generated energy.

   \subsubsection{\bf {Inverters in Power Systems}}
   
The inverter links all DC sources to the AC load, converting energy with over 95\% efficiency. Its size is based on the maximum power of the system and is calculated using Equation (\ref{inv}) \cite{lata2024optimization}.

  \begin{align}
      P_{inv} = \frac{P_{peak}}{\eta_{inv}}\label{inv}
  \end{align}
  
Where $P_{inv}$ - the power of the inverter, $P_{peak}$ is the total maximum load of the system, and
$\eta_{inv}$ - the efficiency of the inverter.\\

\begin{table}[ht]
\centering
\scriptsize
\setlength{\tabcolsep}{3pt} 
\renewcommand{\arraystretch}{1.0} 
\caption{Reliability indicators of Grid-connected hybrid configurations (DG, BG, WP, PV)}
\begin{tabular}{lcccccc}
\hline
\text{System} & \text{\shortstack{WP\\ (kWh/yr)}} & \text{\shortstack{BG \\(kWh)}} & \text{\shortstack{PV\\ (kWh/yr)}} & \text{\shortstack{DG\\ (kWh)}} & \text{\shortstack{Grid \\(kWh)}} & \text{\shortstack{Elec. prod. \\(kWh/yr)}} \\
\hline
BG/Grid                        &        & 525600 &        &        & 3567751 & 4093351 \\
PV/BG/                   &&&&&&\\
Grid/conv                 &        & 525600 & 5251   &        & 3563934 & 4094786 \\
BG/batt/                  &&&&&&\\
Grid/conv               &        & 525600 &        &        & 3567751 & 4093351 \\
PV/BG/batt/               &&&&&&\\
Grid/conv            &        & 525600 & 5239   &        & 3562786 & 4093624 \\
WP/BG/                   &&&&&&\\
Grid/conv                 & 678    & 525600 &        &        & 3567256 & 4093534 \\
PV/WP/BG/                  &&&&&&\\
Grid/conv              & 1016   & 525600 & 12310  &        & 3559398 & 4098325 \\
WP/BG/batt/              &&&&&&\\
Grid/conv            & 1016   & 525600 &        &        & 3566787 & 4093403 \\
PV/WP/BG/               &&&&&&\\
batt/Grid/conv         & 2372   & 525600 & 13154  &        & 3562034 & 4103160 \\
DG/Grid                         &        &        &        & 525600 & 3567751 & 4093351 \\
PV/DG/                     &&&&&&\\
Grid/conv                 &        &        & 5251   & 525600 & 3563934 & 4094786 \\
Grid only                       &        &        &        &        & 4093351 & 4093351 \\
\hline
\end{tabular}
\label{energy}
\end{table}

\begin{table}[ht]
\centering
\scriptsize
\setlength{\tabcolsep}{3pt} 
\renewcommand{\arraystretch}{1.0} 
\caption{Renewable fraction, excess electricity, and reliability indicators of Grid-connected hybrid configurations of (DG, BG, WP, PV)}
\begin{tabular}{lcccc}
\hline
\text{Systems} & \text{\shortstack{Ren \\ Frac (\%)}} & \text{\shortstack{Excess elec.\\ (kWh/yr)}} & \text{\shortstack{Unmet load \\(kWh/yr)}} & \text{\shortstack{MRP\\ (\%)}} \\ 
\hline
BG/Grid & 12.8 & 0 & 0 & 0 \\
PV/BG/      &&&&  \\ 
Grid/conv & 12.9 & 1234 & 0 & 3.16 \\
BG/batt/        &&&&  \\ 
Grid/conv & 12.8 & 0 & 0 & 0 \\
PV/BG/batt          &&&&  \\ 
/Grid/conv & 13.0 & 12.3 & 0 & 3.15 \\
WP/BG/              &&&&  \\ 
Grid/conv & 12.9 & 157 & 0 & 2.04 \\
PV/WP/BG            &&&&  \\ 
/Grid/conv & 13.0 & 4535 & 0 & 7.65 \\
WP/BG/batt          &&&&  \\
/Grid/conv & 12.9 & 1.54 & 0 & 3.06 \\
PV/WP/BG/           &&&&  \\ 
batt/Grid/conv & 13.0 & 9508 & 0 & 9.56 \\
DG/Grid & 0 & 0 & 0 & 0 \\
PV/DG/          &&&&  \\ 
Grid/conv & 0.0932 & 1234 & 0 & 3.16 \\
Grid only & 0 & 0 & 0 & 0 \\
\hline
\end{tabular}
\label{rene}
\end{table}

 The contribution of different energy sources of Grid-connected hybrids presented in Table \ref{energy} and Fig. \ref{scheme} (d) yearly. The Grid remains the dominant supplier, consistently delivering approximately (3.56M-3.57M) kWh per year, which confirms the predominant of Grid dependent rather than fully renewable. Among renewable sources, BG contributes the most significant share, providing a steady 525,600 kWh/yr in all designs that include it. PV and WP contribute comparatively smaller portions, with maximum outputs of about 13154 kWh/yr and 2372 kWh/yr, respectively. The inclusion of DG results in an energy output equivalent to that of the BG system. The total annual generation remains nearly constant at approximately (4.09-4.10)M kWh, underscoring the stabilizing role of the Grid  in balancing supply and demand across various configurations as illustrated in Table \ref{energy}.

\begin{table}[ht]
\label{cos}
\centering
\scriptsize
\setlength{\tabcolsep}{3pt} 
\caption{Cost and energy Production of hybrid renewable energy System (PV,WP,BG \& DG)}
\begin{tabular}{@{}lccccccc@{}} 
\hline
\text{Systems} & \text{\shortstack{COE\\ (\$/kWh)}} & \text{NPC (\$)} & \text{\shortstack{DG\\(kWh/yr)}} & \text{\shortstack{BG\\(kWh/yr)}} & \text{\shortstack{PV\\(kWh/yr)}} & \text{\shortstack{WP\\(kWh/yr)}} & \text{\shortstack{Elec. prod\\(kWh/yr)}} \\
\hline
PV/BG/           &&&&&&& \\
batt/conv & 0.399 & 25.7M &  & 299570 & 11080614 &  & 6958182 \\
PV/DG/              &&&&&&& \\
batt/conv & 0.401 & 25.8M & 299642 &  & 11080614 &  & 6958266 \\
PV/WP/BG/               &&&&&&& \\
batt/conv & 0.407 & 26.2M &  & 297294 & 11588291 & 8082 & 7473806 \\
PV/WP/DG/                   &&&&&&& \\
batt/conv & 0.409 & 26.3M & 297312 &  & 11588291 & 8082 & 7473826 \\
PV/batt                    &&&&&&& \\
/conv & 0.459 & 29.6M &  &  & 12382944 &  & 7893940 \\
PV/WP/                     &&&&&&& \\
batt/conv & 0.466 & 30.0M &  &  & 12567437 & 15153 & 8095730 \\
\hline
\end{tabular}
\label{Tab}
\end{table}

From the comparison in Table \ref{rene}, the addition of PV and WP slightly increases the renewable share from 12.8\% to 13.0\%. However, it also causes more unused energy, reaching about 9508 kWh/yr in the full PV/WP/BG/batt/Grid. All arrangements keep full reliability with no unmet load, but higher renewable use increases the maximum reliability parameter from 0\% to 9.56\%, showing that more Grid is needed to handle solar and wind changes. In contrast, diesel or Grid  have no renewable power or extra energy, but fully reliable, showing a tradeoff between renewable use and system stability.

Table \ref{Tab} presents a comparative assessment of different designs of hybrid energy systems that compromise the components of PV, WP, BG, and DG based on their economic feasibility and performance in energy generation. Designs are evaluated based on two primary cost metrics, NPC and COE as well as their annual electricity output. The energy sources considered include PV, WP, BG, and DG, supported by a battery and converter system. Among the configurations, PV/BG/batt/conv and PV/DG/batt/conv are the most cost effective, exhibiting the lowest COE (\$0.399/kWh and \$0.401/kWh) and NPC (\$25.7 M and \$25.8 M), respectively, as illustrated in Table \ref{Tab} and Fig. \ref{scheme} (c). This indicates that supplementing intermittent PV supply with adjustable sources such as biomass significantly improves the system’s economic performance by increasing the required battery storage and ensuring a more reliable power supply. It shows that a BG can effectively act as a renewable, dispatchable backbone, minimizing battery costs while ensuring reliability and full sustainability. In contrast,  pure renewable systems (PV/batt/conv and PV/ WP/batt/conv) are the most expensive to build and struggle with the highest over generation, making them less economically efficient.

\subsection{\bf {Economic Assessment of Hybrid Energy Systems}}

The NPC is the total lifetime cost of a hybrid system, covering capital, maintenance, replacements, fuel, emissions, and grid use, adjusted for inflation and interest to reflect the current-dollar value \cite{emezirinwune2024off,MULENGA2023601}. The NPC is the current value of the installation and operating costs of a system over its lifetime. HOMER computes it using (\ref{crf}), subtracting the present value of all income, including power sales, and the salvage value is given in equation (\ref{crf}) \cite{ennemiri2024optimization,ABID2021120656}.

\begin{align}
    NPC_{total}=\frac{C_{ann,tot}}{CRF} \label{crf}
\end{align}

Where CRF - the capital recovery factor, $NPC_{total}$ - the total net present cost and C\textsubscript{ann}, \textsubscript{tot} is the total annualized costs of all components. HOMER uses it to compute the LCOE and the net present value. COE - the average cost per kWh of electricity, found by dividing the annualized cost by total electricity, as shown in (\ref{tot}) \cite{emezirinwune2024off,ODOU20201266}. 

\begin{align}
    COE=\frac{C_{ann,tot}}{E_{t}} \label{tot}
\end{align}

Where $COE$ - the cost of energy, $E_{t}$ - the total energy produced over time. 
According to \cite{emezirinwune2024off,mukhtar2024enviro} the capital recovery factor is calculated from the interest rate (i) and the project life (n), as given in equation (\ref{n}).

In a hybrid PV/wind/ biomass/ diesel/ batteries system, the annualized cost is calculated using equation (\ref{co}).

\begin{align}
    C_{ann,tot} & =\sum_{N=1}^{N_{PV}} C_{ann,PV } + \sum_{N=1}^{N_{WP}} C_{ann,WP } \nonumber \\
      & \quad + \sum_{N=1}^{N_{BG}} C_{ann,Bio} + \sum_{N=1}^{N_{Dsl}} C_{ann,DG }  \nonumber \\
      & \quad + \sum_{N=1}^{N_{batt}} C_{ann,batt}  + \sum_{N=1}^{N_{conv}} C_{ann,conv} \label{co}
\end{align}
   
Where $C_{ann,tot}$ - the total annual cost, $C_{ann, PV}$ - the annual cost of PV modules, $C_{ann,WP}$ - the annual cost of wind turbines, $C_{ann, BG}$ - the annual cost of the biomass gasifier, $C_{ann,DG}$ - the annual cost of diesel generators, \textsubscript{batt} - the annual cost of batteries, and $C_{ann,conv}$ - the annual cost of the converter. $N_{PV}$ - the number of PV modules, $N_{WP}$ - the number of wind turbines, N\textsubscript{BG} - the number of biomass gasifiers, $N_{DG}$ - the number of diesel generators, $N_{batt}$ - the number of batteries, and $N_{conv}$ - the number of converters \cite{emezirinwune2024off}.

\begin{align}
    CRF=\frac{i(i+1)^n}{(1+i)^n-1} \label{n}
\end{align}

\begin{table}[ht]
\centering
\scriptsize
\setlength{\tabcolsep}{2.5pt} 
\renewcommand{\arraystretch}{1.0} 
\caption{Cost and performance summary of different PV, WP, BG and DG hybrid systems}
\begin{tabular}{lccccccc}
\hline
\text{Systems} & \text{\shortstack{Initial\\cost (\$)}} & \text{\shortstack{Oper.\\cost (\$)}} & \text{\shortstack {O \& M \\(\$/yr)}} & \text{\shortstack{Ren.\\Frac (\%)}} & \text{\shortstack{Excess elec.\\(kWh/yr)}} & \text{\shortstack{Unmet load\\ (kWh/yr)}} & \text{\shortstack {MRP\\ (\%)}} \\
\hline
PV/BG/                          &&&&&&& \\
batt/conv        & 20.6M & 328258 & 211129 & 100  & 8671857 & 2867 & 6625 \\
PV/DG/                      &&&&&&& \\
batt/conv        & 20.5M & 337121 & 209123 & 92.7 & 8671938 & 2867 & 6625 \\
PV/WP/                      &&&&&&& \\
BG/batt/conv     & 21.1M & 322854 & 203605 & 100  & 9613875 & 2751 & 6931 \\
PV/WP/                          &&&&&&& \\
DG/batt/conv     & 21.1M & 331649 & 201613 & 92.7 & 9613048 & 2751 & 6931 \\
PV/batt/                            &&&&&&& \\
conv           & 23.3M & 398660 & 244862 & 100  & 8778048 & 2907 & 7404 \\
PV/WP/                              &&&&&&& \\
batt/conv        & 23.7M & 403323 & 240603 & 100  & 9088279 & 2969 & 7519 \\
\hline
\end{tabular}
\label{cost1}
\end{table}

The comparative assessment shows that hybrid systems that include BG deliver the strongest techno-economic performance, achieving a full renewable fraction alongside lower operating costs and reduced O and M requirements. Among these, the PV/WP/BG/batt/conv configuration  performs optimally (see  Fig. \ref{scheme} (b)), offering low operating costs, minimal unmet load, and substantial excess electricity due to the strong complementarity of solar and wind resources. As shown in   Fig. \ref{scheme} (a), diesel-supported systems despite slightly lower initial costs exhibit higher operating expenses and reduced renewable penetration, reflecting continued dependence on fossil fuels. Fully renewable systems without BG incur the highest costs and unmet load, highlighting the value of dispatchable BG support to improve both economic viability and operational reliability, as illustrated in Table \ref{cost1}.

The economic analysis in Table \ref{per} emphasizes clear tendencies and trade-offs among hybrid energy systems. BG based design outperforms the DG and Grid only options, showing lower levelized COE (\$0.097/kWh-\$0.0983/kWh) and NPC (\$6.25M-\$6.34M) due to reduced fuel and O\&M expenses, despite higher capital costs. Adding PV, DG or battery components slightly raises COE and NPC while improving the renewable fraction, reflecting a modest cost advantage for higher sustainability. In the BG-WP-PV-DG-batt-conv configuration, the components, the inclusion of the DG marginally increased the emission levels but provided critical reliability during periods of low renewable energy output, such as during cloudy covers. The small increase in pollutants as a result of the limited use of diesel.

\begin{table}[ht]
\centering
\scriptsize
\setlength{\tabcolsep}{2.5pt} 
\renewcommand{\arraystretch}{1.0} 
\caption{Comparative economic performance of the Grid with different hybrid energy systems (DG, BG, WP, PV)}
\begin{tabular}{lccccccc}
\hline
\text{Systems} & \text{\shortstack{COE \\(\$/kWh)}} & \text{\shortstack{NPC\\ (\$)}} & \text{\shortstack{Initial \\cost (\$)}} & \text{\shortstack{Oper. cost \\(\$)}} & \text{\shortstack{O \& M\\ (\$/yr)}} \\
\hline
BG/Grid                        & 0.0970 & 6.25M & 60000   & 0.00   & 393131  \\
PV/BG/                  &&&&&   \\
Grid/conv                 & 0.0970 & 6.26M & 67656   & 0.00   & 392788   \\
BG/batt/                        &&&&& \\
Grid/conv               & 0.0970 & 6.26M & 63169   & 0.00   & 393218   \\
PV/BG/batt/                     &&&&& \\
Grid/conv            & 0.0971 & 6.26M & 73582   & 0.00   & 392837   \\
WP/BG/                              &&&&& \\
Grid/conv                 & 0.0973 & 6.27M & 74098   & 0.00   & 393381  \\
PV/WP/BG/                               &&&&& \\
Grid/conv              & 0.0975 & 6.29M & 98795   & 0.00   & 392829   \\
WP/BG/batt/                        &&&&&  \\
Grid/conv            & 0.0975 & 6.29M & 87312   & 0.00   & 393654  \\
PV/WP/BG/                           &&&&& \\
batt/Grid/conv         & 0.0983 & 6.34M & 133946  & 0.00   & 393856  \\
DG/Grid                         & 0.0999 & 6.44M & 21500   & 34690  & 407759    \\
PV/DG/                              &&&&& \\
Grid/conv                 & 0.1000 & 6.45M & 29156   & 34690  & 407417   \\
Grid only                       & 0.1000 & 6.45M & 0.00    & 0.00   & 409335     \\
\hline
\end{tabular}
\label{per}
\end{table}

\begin{table}[ht]
\centering
\scriptsize
\setlength{\tabcolsep}{2pt} 
\caption{Design, component sizing, energy production, and associated emissions of hybrid renewable energy systems} \label{size}
\begin{tabular}{@{}llllll@{}} 
\hline
\text{Design} & \text{Components} & \text{Name} & \text{Size} & \text{\shortstack{Production\\(kWh/yr)}} & \text{\shortstack{Emission\\(kg/yr)}}\\
\hline
 & PV & plate PV & 6,323.0 kW & 11,093,773 &  \\
BG/WP/PV/\\
conv/batt & ASM & 14,375.0 strings &  &  & \\
 & converter & 1401.0 kW &  &  & \shortstack{$CO_{2}$= 73.00 \\ CO=7.52 \\ SO$_2$= 0.00 \\ NO$_x$=67.10}\\
\hline
BG/WP/PV/\\
DG/conv/batt & Generator & BG & 60.0 kW & 285,896 & \\
 & PV & plate PV & 6,323.0 kW & 11,093,773 & \\
 & Storage & ASM & 14,375.0 strings &  & \\
 & converter & 1401.0 kW &  &  & \shortstack{$CO_{2}$=260,602.00 \\ CO=643.00            \\ SO$_2$= 523.00            \\ NO$_x$=5740.00}\\
\hline
BG/PV/\\
conv/batt & PV & plate PV & 7,058.0 kW & 11,380,192 & \\
 & Storage & ASM & 17,428.0 strings &  & \\
 & converter & 1,473.0 kW &  &  & free\\
\hline
DG/PV/WP/\\
conv/batt & Generator & DG & 60.0 kW & 299,644 & \\
 & PV & plate PV & 6,316.0 kW & 11,080,614 & \\
 & Storage & ASM & 14,295.0 strings &  & \\
 & converter & 1,439.0 kW &  &  & \shortstack{$CO_{2}$=260781.00          \\ CO= 644.00  \\ SO$_2$=524.00  \\ NO$_x$=5744.00}\\
\hline
BG/WP/PV/DG/\\
Grid/conv/batt & Generator & BG & 60.0 kW & \shortstack{purchases\\(3,563,933)} & \\
 & PV & plate PV & 2.91 kW & 5,251 & \\
 & Converter & converter & 1.23 kW & 525,600 & \\
 & Grid & Grid & 999,999.0 kW &  & \shortstack{$CO_{2}$=2252534.00 \\ CO= 13.20  \\ SO$_2$=9765.00 \\ NO$_x$=4893.00}\\
\hline
\end{tabular}
\end{table}

\subsection{Discussion}

In this paper, the biogas generator (eq. \ref{ebm}) provides a stable base load, while solar PV panels (eq. \ref{pv}) and wind turbines (eq. \ref{pg3}) supplement the generation under favorable conditions. The DG (eq. \ref{dsl}) operates as a backup during periods of low renewable output. The ASM battery (eq. \ref{batt} and \ref{inv}) strengthens the  reliability of the system by storing excess energy and mitigating power variations. This design reduces fuel dependence and boosts system resilience, but it also requires higher capital investment and includes more complex multi-source integration. Therefore, battery storage improves reliability by sustaining power supply during low solar irradiance. In addition, the hybrid design option confirms the continuous operation of the Boru Meda Hospital with a  high renewable allocation and a reduction in emissions.
 In the design of the system, the BG provides a continuous and constant power delivery and PV energy during the sunny season. 

 Table \ref{size} shows hybrid configurations that combine BG, WP, PV, DG, converters, batteries, and in some cases, Grid connections. Each setup is evaluated for component sizing, annual energy output, and emissions. The BG/WP/PV/conv/batt system offers a good balance of reliability and sustainability, taking advantage of the available solar and biomass resources. Integration of biomass and wind increases stability and compensates for solar variability, while BG provides backup power with minimal emissions.
Systems that integrate DG, PV, WP, BG, converters and batteries (DG/WP/BG/PV/conv/batt) exhibit the highest present value (\$307,113) and annual value (\$19,496), indicating strong economic viability, as shown in Table \ref{hyb}. In contrast, purely renewable systems such as BG/WP/PV/conv/batt show lower present and annual worth (\$176,452 and \$11,202) due to limited dispatchable generation. DG-inclusive configurations achieve high ROI ($\approx$ 20\%) and IRR ($\approx$ 18\%) by providing reliable energy output, while renewable-only systems have very low ROI (0.5\%) and IRR (1.4\%), reflecting economic challenges under intermittent solar and wind conditions. Simple and discounted repayment periods for DG-integrated systems are short (7.21–8.71 years), while renewable-dominant systems require 16–17 years, indicating slower capital recovery (Table \ref{hyb}). In general, hybrids combining DG and renewable offer both financial attractiveness and reliability, whereas purely renewable configurations improve sustainability but may require policy support to improve economic feasibility. System selection should balance economic performance, energy reliability, and environmental benefits, especially for hospital or community applications.
 
 \begin{figure}[ht]
     \centering
\includegraphics[scale = 0.45]{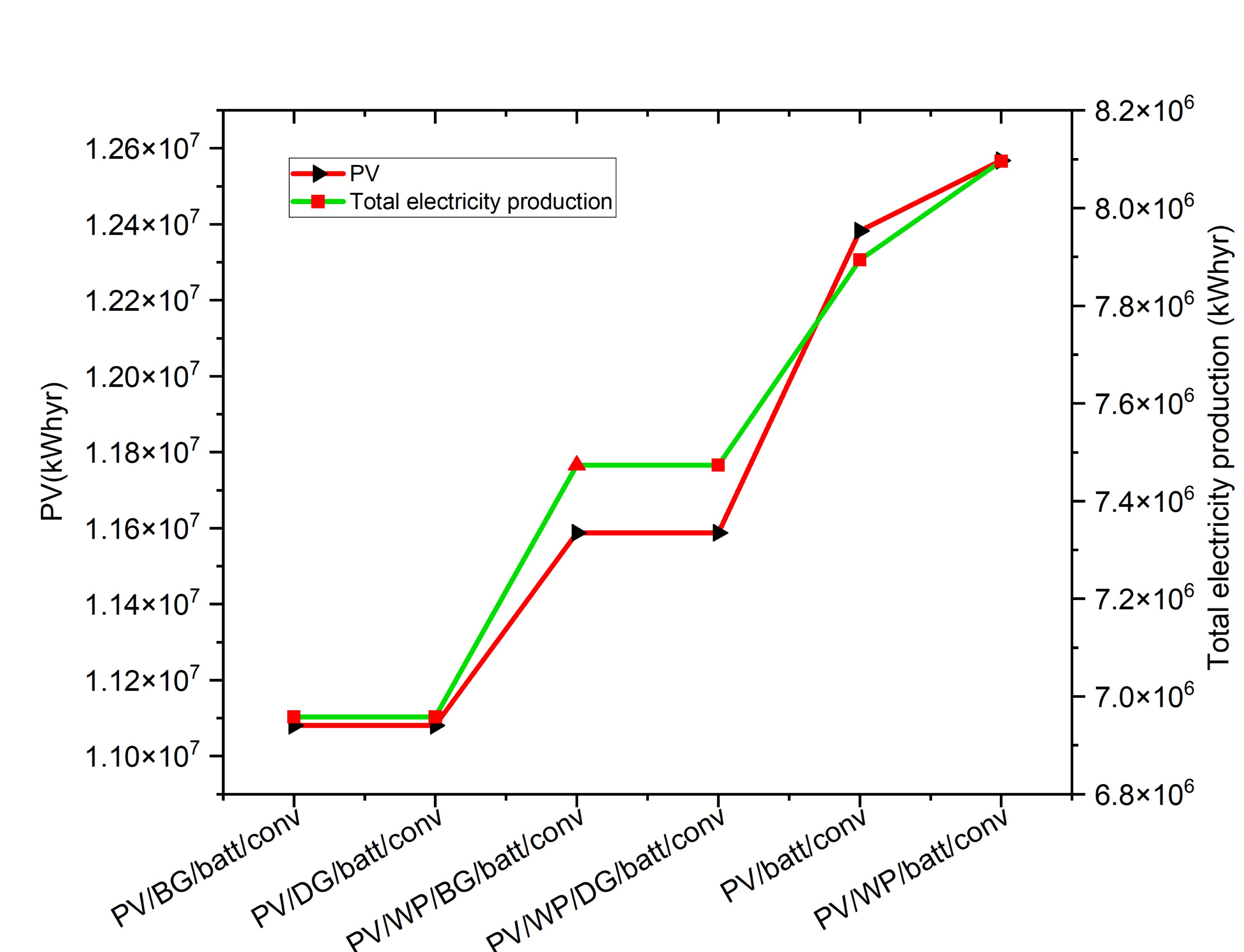}
     \caption{Comparison of annual PV energy and total energy production}\label{pv1} 
   \end{figure}

   Design performance shows a clear positive correlation with PV integration as evidenced by DG/BG hybrid toward PV and PV/WP dominated setups. The apex configuration PV/WP/batt/conv capitalizes on solar's diural abundance, wind's temporal complementarity, and storage driven  Grid stability to gain a maximum annual yield and system reliability. Hence, solar energy, as the primary contributor within a hybrid energy setup, exceeds the convectional (diesel, biomass) and variable like wind power alternatives a net energy yield as illustrated in Fig. \ref{pv1}.
   
The high potential for solar energy, combined with the modest yet usable wind resource, provides a strong basis to gain operational efficiency and improve the economic competitiveness of the presented energy system for the Hospital. As presented in Fig. \ref{pv1}, the PV/BG/batt/conv configuration appears to be the most balanced and optimal system in all evaluation criteria, particularly in COE and NPC. This design achieves a low initial capital cost and the lowest NPC while delivering a 100\% renewable fraction, high reliability (minimal unmet load) and reasonable operating and maintenance costs. The results show that incorporating a BG as a renewable and adjustable component provides a stable backbone for the system, effectively reducing the battery capacity requirements while maintaining reliability and ensuring full sustainability. In contrast, the purely renewable alternatives PV/batt/conv and PV/WP/batt/conv exhibit the highest capital and life cycle costs and suffer from notable excess power production. These limitations make them less economically efficient despite their full reliance on renewable resources.

   \begin{figure}[ht]
    \centering
    \hspace{-0.5cm} 
\includegraphics[scale=0.45]{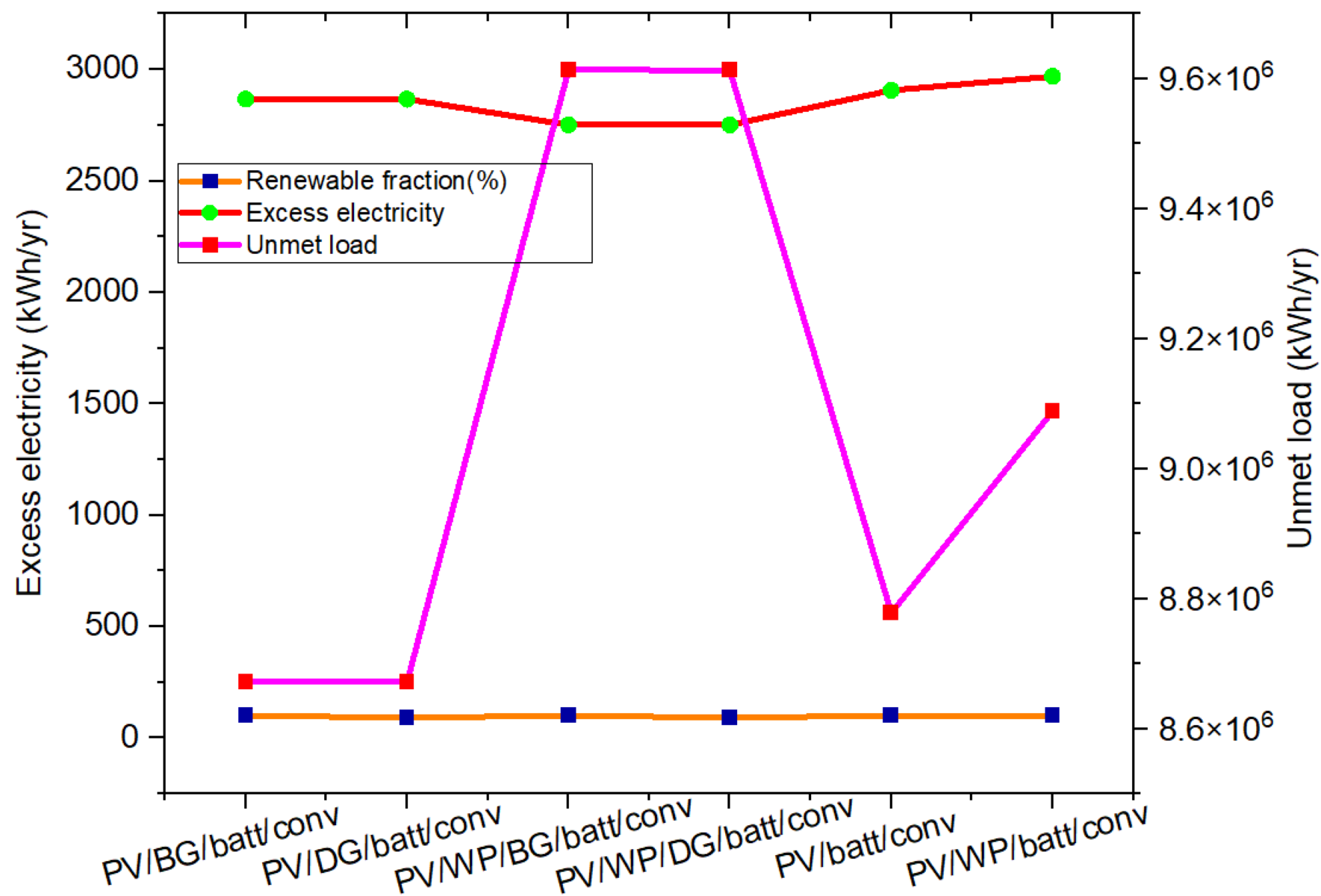}
    \caption{Comparative performance of hybrid energy system configurations (RF, excess electricity and unmet load)} \label{grid}
\end{figure}
As illustrated in Fig. \ref{grid}, a comparative performance assessment of different hybrid energy systems in terms of  excess electricity generation, renewable fraction, and unmet load showing clear trade-offs among system designs. Setups including multiple renewable sources and storage conversion units exhibit consistently high renewable fractions, indicating a strong contribution of renewable to total energy supply across all cases. Excess electricity remains relatively high and stable, particularly for systems with multiple production like PV/WP/batt/conv combinations, which enhances consistency but may also indicate opportunities for improved load matching or energy utilization strategies. In contrast unmet load varies more significantly among designs, with systems lacking sufficient generation diversity or flexibility showing pronounced peaks, highlighting their reduced ability to satisfy demand continuously. Generally, the results underscore the importance of integrating complementary renewable resources and adequate storage conversion infrastructure to minimize unmet load while maintaining a high renewable fraction, thereby achieving a balanced compromise between system reliability, energy efficiency, and sustainable operation.

 Grid interconnection substantially improves system reliability by enabling bidirectional power exchange allowing electricity imports during supply deficits and exports of extra renewable energy to the Grid. Hybridization of renewable and conventional energy sources ensures continuous operation and enables dynamic dispatch optimization, effectively balancing cost, emissions, and fuel consumption as shown in Table \ref{size}. Although the inclusion of Grid and diesel components slightly increases total emissions, the high share of renewable energy significantly offsets the environmental footprint, making this configuration a technically robust and environmentally sustainable solution for critical load applications. This shows that supplementing intermittent PV supply with adjustable sources such as biomass significantly improves the system’s economic performance by improving the required battery storage and ensuring a more reliable power supply.

   \begin{figure}[ht]
    \centering
\includegraphics[scale=0.45]{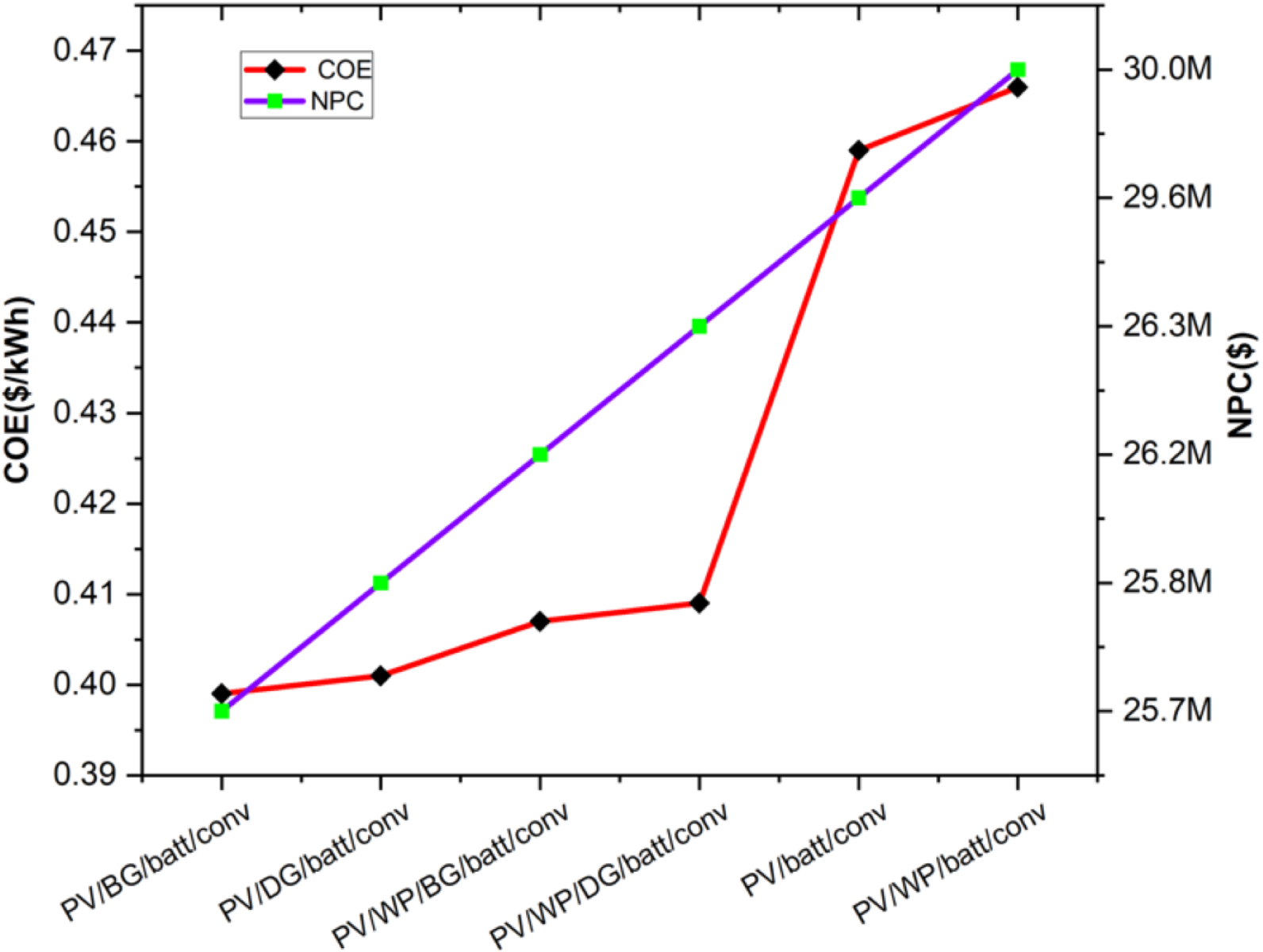}
    \caption{Economic cost, net present cost, and energy output of PV-WP-BG hybrid system}
    \label{value}
\end{figure}

\begin{table}[ht]
\centering
\scriptsize
\setlength{\tabcolsep}{2.5pt} 
\renewcommand{\arraystretch}{1.0} 
\caption{Economic performance indicators of different hybrid system configurations}
\begin{tabular}{lccccccc}
\hline
\text{Designs} & \text{\shortstack{present \\ worth(\$)}} & \text{\shortstack{Annual\\ worth(\$)}} & \text{\shortstack{Return on \\investment (\%)}} & \text{\shortstack{Internal rate\\ of return(\%)}} & \text{\shortstack{simple pay\\ back (yr)}} & \text{\shortstack{Discounted \\pay back(yr)}}\\
\hline
DG/WP/BG/\\
PV/conv/batt & 307,113 & 19,496 & 20.00 & 18.40 & 7.21 & 8.64\\
BG/WP/\\
PV/conv/batt & 176,452 & 11202 & 0.50 & 1.40 & 16.40& 16.05\\
BG/PV\\
/conv/batt & 306,313 & 19,445 & 20.00 & 18.30 & 7.26& 8.71\\
BG/PV/WP/DG/\\
Grid/conv/batt & 174,576 & 11,083 & 0.50 & 1.40 & 16.39 & 17.00\\
\hline
\end{tabular}
\label{hyb}
\end{table}

Although diesel generator systems (PV/DG/batt/conv and PV/WP/DG/batt/conv) may initially be cost competitive, they promise operators to use fossil fuels and cannot achieve a fully renewable energy target, making them less sustainable in the long term (see  Fig. \ref{pv1}). The BG/PV/conv/batt system, powered by solar and biomass, provided clean, emission free, and reliable energy supporting the findings of Ennemiri \textit{et al.} \cite {ennemiri2024optimization}. The slightly larger PV capacity relative to other configurations enhanced system stability and reduced dependence on fossil-based backup generation. This configuration is one of the most sustainable and self-sufficient designs, particularly suitable for regions such as Boru Meda with high solar irradiance and readily available biomass resources. Therefore, the PV-BG hybrid system offers the strongest balance between economic and environmental goals (see Fig. \ref{value}), consistent with the results reported in \cite{motevakel2025strategic,habib2025multi}.

The comparative assessment further highlights the Grid offsetting potential of battery storage. In systems incorporating batteries, such as PV/BG/batt/Grid/conv compared to PV/BG/Grid/ conv, the Grid input decreases marginality (from 3,563,934 kWh to 3,562,786 kWh) as shown in Table \ref{energy}, while the load output remains nearly constant. This shows that batteries facilitated limited renewable energy utilization during peak demand, slightly reducing Grid dependence without substantially increasing total renewable penetration. The most integrated design PV/WP/BG/batt/conv achieves the highest total generation (4,103,160 kWh/yr) and the most diversified renewable mix, yet still relies on the Grid for about 87\% of its supply. These results emphasize that, although hybrid systems enhance renewable integration, Grid dependency remains predominant, with batteries providing incremental efficiency rather than full energy independence.

Compared to the DG/BG/Grid/WP/PV/batt/conv design, BG/WP/PV/batt/conv achieved substantial emission decreases of 99.99\% for $CO_{2}$, 43.03\% for CO, 100\% for $SO_{2}$, and 98.63\% for $NO_{2}$. In disparity, BG/PV/DG/WP/batt/conv noted declines of 88.43\% in $CO_{2}$, 94.64\% in $SO_{2}$, but showed a significant increase of 97.95\% in CO and 14.76\% in $NO_{x}$ emissions relative to the baseline system, mainly due to incomplete biomass combustion. Similarly, the configuration DG/BG/WP/PV/batt/conv caused remarkable reductions of 88.42\% in $CO_{2}$ and 94.64\% in $SO_{2}$; however, it exhibited substantial increases of 97.95\% in CO and 14.76\% in $NO_{x}$ emissions as a result of incomplete combustion and transient high temperature peaks, respectively. These comparative results, summarized in Table \ref{size}, highlight the superior environmental performance of the BG/PV/batt/conv design.

The BG-WP-PV-DG-Grid-conv-batt, which results in high $CO_{2}$ emissions from carbon-intensive electricity supplied by the Grid, performs poorly in environmental terms. However, the system ensures high reliability and energy security. The result indicates the critical role of expanding the penetration of renewable energy in Grid-connected systems to minimize emissions and enhance long-term energy sustainability, consistent with the conclusions reported in \cite{ba2025optimal}.

\section{Conclusion}

In this study, HOMER pro 3.11.2 was used to evaluate the detailed techno-economic performance of more than 40,000 hybrid designs, with approximately 1000 feasible HES setups simulated for Boru Meda Hospital. The Scenario evaluation showed that interconnection of PV, BG, WP, DG, converter, and battery components can effectively provide the hospital’s daily energy consumption of 11,214.66 kWh while conforming reliability and reducing emissions. Among the designs examined, the PV/BG/batt/conv system is identified as the most economically viable and environmentally friendly approach. It achieved the lowest COE (\$0.339/kWh) and NPC (\$25.7 M), coupled with a 100\%  renewable fraction, minimal unmet load and reasonable O \& M costs. The Integration of biomass as a flexible clear energy source significantly improved the reliability of the system and reduced the need for large battery storage or diesel backup, making it a robust and carbon free option.Although adding wind power marginally improved total power generation, it also addressed the overall cost of the system, indicating that wind integration is less economically favorable under current local conditions.

However, systems heavily reliant on diesel or Grid power, although reliable, showed high emissions and poor sustainability performance, pointing out the need to prioritize renewable-based set ups. DG-integrated hybrids exhibit advanced techno-economic capability with significant value, strong ROI (~20\%) and IRR (~18\%), backed by fast capital recovery (7.21–8.71 years). In contrast, only renewable systems show low value, minimal ROI (~0.5\%), and long payback periods (16–17 years) due to wind-solar variability.
 
Ultimately, the PV/BG hybrid represents the Optimal system design for Boru Meda Hospital, balancing cost, reliability, and sustainability. It facilitates a scale energy, resilient, and low emission solution suitable for rural health facilities in Ethiopia, proving the potential of hybrid renewable systems to replace carbon intensive generation  and improve energy security in significant public entities.

\section{Limitations and Future Work}

This article is limited to simulation based analysis using HOMER Pro simulations and does not include physical prototypes, field measurements, or real-time operational data. As a result, many practical aspects, such as installation challenges, environmental fluctuation, and long-term system decline, remain unexamined. Furthermore, the evaluation does not incorporate in depth biomass characterization or logistics assessments. These limitations will be addressed in future research that should incorporate real time data collection, prototype development, and on site performance analysis to verify and improve the simulation outcomes. Further studies could also analyze economic practicability and long term resilience to upgrade the potential of the system for real world application.

   \textbf{CRediT authorship contribution statement}

    \textbf{Tegenu Argaw Woldegiyorgis}  led the conceptual framework, methodological design, and data analysis, and authored the initial draft while overseeing the review and revision stages \textbf{Hong Xian Li} contributed supervision and assisted in the review and revision of the manuscript. \textbf{Eninges Asmare} managed and analyzed the data, verified the findings and improved the manuscript, \textbf{Fekadu Chekol Admassu} Evaluated the results and contributed to the manuscript’s revision. \textbf{Merkebu Gezahegne} collected data and contributed to writing and reviewing. \textbf{Abdurehman Kebede} led the data collection and supported the writing and review of the manuscript. \textbf {Haris Ishaq} visualized the results and contributed to reviewing, revising and editing the manuscript. \textbf {Tadese Abera} contributed to the collection of data and reviewed the manuscript. 

\textbf{Declaration of competing interest}

The authors declare that they have no known competing financial interests or personal relationships that could have appeared to influence the work reported in this paper.

\textbf{Data Availability Statement}

Data supporting the findings of this study are available from the corresponding author on a reasonable request.

\textbf{Conflict of interest}

The authors have no conflicts of interest to announce.

\textbf{Declaration of competing interest}

The authors declare that they have no competing financial interests that could have been performed to influence this work.

\textbf{Funding}

This research article did not receive a specific grant from any funding agency.

\bibliography{sn-bibliography}

\end{document}